%% file: SCSS-2021.tex
\documentclass[submission,copyright,creativecommons]{eptcs}
\providecommand{\event}{SCSS 2021} % Name of the event you are submitting to
\usepackage{breakurl}             % Not needed if you use pdflatex only.
\usepackage{underscore}           % Only needed if you use pdflatex.
\usepackage{fancyhdr}
\usepackage{xcolor}
\usepackage{graphicx}
\usepackage{amsmath}
\usepackage{epstopdf}
\usepackage{caption}
\usepackage{subcaption}
\usepackage{threeparttable}
\usepackage{array,multirow,graphicx}
\usepackage{colortbl}
\usepackage{algorithm}
\usepackage{algpseudocode}

\newcommand{\Mbarka}[1]{{{ #1}}}

\newcommand{\eg}{e.g.,}

\title{Failure Analysis of Hadoop Schedulers using an Integration of Model Checking and Simulation}
\author{Mbarka Soualhia
\institute{Concordia University,\\  Montr\'eal, Canada}
\email{soualhia@ece.concordia.ca}
\and
Foutse Khomh  
\institute{Polytechnique Montr\'eal,\\ Montr\'eal, Canada}
\email{\quad foutse.khomh@polymtl.ca }
\and
Sofi\`ene~Tahar
\institute{Concordia University,\\  Montr\'eal, Canada}
\email{\quad tahar@ece.concordia.ca }
}
\def\titlerunning{Failure Analysis of Hadoop Schedulers using an Integration of Model Checking and Simulation}
\def\authorrunning{M. Soualhia, F. Khomh \& S. Tahar}
\begin{document}
\sloppy

\maketitle

\begin{abstract}
\textbf{Abstract:} The Hadoop scheduler is a centerpiece of Hadoop, the leading processing framework for data-intensive applications in the cloud.
Given the impact of failures on the performance of applications running on Hadoop, testing and verifying the performance of the Hadoop scheduler is critical.
Existing approaches such as performance simulation and analytical modeling are inadequate because they are not able to ascertain a complete verification of a Hadoop scheduler. This is due to the wide range of constraints and aspects involved in Hadoop.
\Mbarka{In this paper, we propose a novel methodology that integrates and combines simulation and model checking techniques to perform a formal verification of Hadoop schedulers, focusing on the following properties: schedulability, fairness and resources-deadlock freeness.} We use the CSP language to formally describe a Hadoop scheduler, and the PAT model checker to verify its properties.
\Mbarka{Next, we use the proposed formal model to analyze the scheduler of OpenCloud, a Hadoop-based cluster that simulates the Hadoop load, in order to illustrate the usability and benefits of our work.}
Results show that our proposed methodology can help identify several tasks failures (up to 78\%) early on, i.e., before the tasks are executed on the cluster.
\end{abstract}
%\keywords{Formal Verification, Model Checking, Simulation, Hadoop Schedulers, OpenCloud Failure Analysis, PAT, CSP.}

\input{introduction}

\input{preliminaries}

\input{framework}

\input{casestudy}
\input{relatedwork}
\input{conclusion}

%\nocite{*}
\bibliographystyle{eptcs}
\bibliography{mabiblio}
\end{document}

%% file: introduction.tex
\section{Introduction}
\vspace{-5pt}
Motivated by the reasonable prices and the good quality of cloud services, several enterprises and governments are deploying their applications in the cloud. Hadoop~\cite{HadoopOnline} has become enormously popular for processing data-intensive applications in the cloud. A core constituent of Hadoop is the scheduler. The Hadoop scheduler is responsible for the assignment of tasks belonging to applications across available computing resources. Scheduling data-intensive applications' tasks is a crucial problem, especially in the case of real-time systems, where several constraints should be satisfied.
Indeed, an effective scheduler must assign tasks to meet applications' specified deadlines and to ensure their successful completions. %using the assigned resources.
In the cloud, failures are the norm rather than the exception and they often impact the performance of tasks running in Hadoop. Consequently, the Hadoop scheduler may fail to achieve its typical goal for different reasons such as resources-deadlock, task starvation, deadline non-satisfaction and data loss~\cite{WOHA2014}~\cite{RAFT2011}.
This results in extra delays that can be added and propagated to the overall completion time of a task, which could lead to resources wastage, and hence significantly decreasing the performance of the applications and increasing failure rates.
%\Foutse{i would removed the following paragraph or rephrased it a bit ....we are making too much emphasis on the failures in the cloud but not much on the failures of the tasks which may be due to poor scheduling...}
For instance, Dinu and Ng~\cite{Dinu:2012} analyzed the performance of the Hadoop framework under different types of failures. They reported that the Hadoop scheduler experiences several failures due to the lack of information sharing about its environment (\eg{} scheduler constraints, resources availability), which can lead to poor scheduling decisions.
For instance, they claim that the Hadoop scheduler experiences several failures because of prioritized executions of some long tasks before small ones, long delays of the struggling tasks or unexpected resources contentions~\cite{Dinu:2012}~\cite{SOUALHIA2017}. 
%\Foutse{in the paragraph below you sound like if the verification is only important to make hadoop scheduler more reliable and you don't emphasize enough the fact that your framework will allow people to simulate the behavior of their clusters and the fact hat they will be able to prevent certain failures from occurring in the field....because they understood the circumstances that would caused them!}
The widespread use of the Hadoop framework in safety and critical applications, such as healthcare~\cite{Real-Time2017} and aeronautics~\cite{NASA}, poses a real challenge to software engineers for the designing and testing of such framework to meet its typical goal and avoid poor scheduling decisions.
Although several fault-tolerance mechanisms are integrated in Hadoop to overcome and recover from these failures, several failures still occur when scheduling and executing tasks~\cite{SOUALHIA2017}.
%\Foutse{which failures? those from the environment or those stemming from poor scheduling?}
Indeed, these failures can occur because of poor scheduling decisions (\eg{} resources deadlock, task starvation) or constraints related to the environment where they are executed (\eg{} data loss, network congestion).
As such, verifying the design of Hadoop scheduler is an important and open challenge to identify the circumstances leading to task failures and performance degradation %Also, it can help check the main issues that can occur while scheduling tasks and their impact on tasks' executions.
%\Foutse{do you want to say that they can anticipate on the potential issues?}
so that software engineers studying Hadoop can anticipate these potential issues and propose solutions to overcome them. This is to improve the performance of the Hadoop framework.\\
\indent Different approaches have been adopted by the research community to verify and validate the behavior of Hadoop with respect to scheduling requirements.
Traditionally, simulation and analytical modeling have been widely used for this purpose. However, with many Hadoop-nodes being deployed to accommodate the increasing number of demands, they are inadequate because they are not efficient 
%\Foutse{do you mean that they do not scale?} \red{I do not like the work scale :(}
in exploring large clusters.
In addition, given the highly dynamic nature of Hadoop systems and the complexity of its scheduler, they cannot provide a clear understanding and exhaustive coverage of the Hadoop system especially when failures occur. %\Foutse{why? can you explain more?}.
Particularly, they are not able to ascertain a complete verification of the Hadoop scheduler because of the wide range of constraints and unpredictable aspects involved in the Hadoop model and its environment (\eg{} diversity of running loads, availability of resources and dependencies between tasks).
%\Foutse{can we be more specific?}
%On the other hand,
Formal methods have recently been used to model and verify Hadoop. However, to the best of our knowledge %Despite many successful results~\cite{Su2009,Towards-Reddy2013,OnoCoq2011} have been published in this direction,
very few efforts %, to the best of our knowledge,
have been invested in applying formal methods to verify the performance of Hadoop (\eg{ data locality, read and write operations, correctness of running application on Hadoop})~\cite{Su2009}~\cite{Towards-Reddy2013}~\cite{OnoCoq2011}.
%\Foutse{what kind of verification did they do then...which is unrelated to behavior...you are trying to say that they didn't verified the behavior of the scheduler?} 
In addition, there is no work that addresses the formal verification of Hadoop schedulers. %so far.
Considering the above facts, we believe that it is important to apply formal methods to verify the behavior of Hadoop schedulers. This is because formal methods are more able to model complex systems and thoroughly analyze their behavior than empirical testing.
Our aim is to formally analyze the impact of the scheduling decisions on the failures rate and avoid simulation cost in terms of execution time and hardware cost (especially for large clusters). This allows to early identify circumstances leading to potential failures, in a shorter time compared to simulation, and prevent their occurrences.
In contrast to simulation and analytical modeling, knowing these circumstances upfront would help practitioners better select the cluster settings (\eg{} number of available resources, type of scheduler) and adjust the scheduler design (\eg{} number of queues, priority handling strategies, failures recovery mechanisms) to prevent poor scheduling decisions in Hadoop.  \\
%\Foutse{why? its not because it havent been done that it should be done...we should provide a better motivation, explaining why it should work intuitively...}
\indent \Mbarka{In this paper,
%we investigate the impact of the Hadoop scheduler performance \Foutse{what do you mean by performance exactly? is it really what you are verifying?} on the failures rate and possible strategies to overcome them. Specifically,
we present a novel methodology that combines simulation and model checking techniques to perform a formal analysis of Hadoop schedulers. Particularly, we study the feasibility of integrating model checking techniques and simulation to help in formally verifying some of the functional scheduling properties
%\Foutse{you said before that you were verifying performance..which is a very wide concept....please be precise and use a consistent terminology!!!}
in Hadoop including schedulability, resources-deadlock freeness, and fairness~\cite{Cheng2002}.}
To this aim, we first present a formal model to construct and analyze the three mentioned properties within the Hadoop scheduler. 
We use the Communicating Sequential Processes (CSP) language~\cite{CSP} to model the existing Hadoop schedulers (FIFO, Fair and Capacity schedulers~\cite{SOUALHIA2017}), and use the Process Analysis Toolkit (PAT) model checker~\cite{PAT} to verify the schedulers' properties.
Indeed, the CSP language has been successfully used to model the behavior of synchronous and parallel components in different distributed systems~\cite{CSP}. Furthermore, PAT is a CSP-based model checker that has been widely used to simulate and verify concurrent, real-time systems, etc.~\cite{PAT}. 
\Mbarka{Based on the generated verification results in PAT, we explore the relation between the scheduling decisions and the failures rate while simulating different load scenarios running on the Hadoop cluster. This is in order to propose possible scheduling strategies to reduce the number of failed tasks and improve the overall cluster performance (resources utilization, total completion time, etc).} 
In order to illustrate the usability and benefits of our work to identify failures that could have been prevented using our formal analysis methodology, we apply our approach on the scheduler of OpenCloud, a Hadoop-based cluster that simulates the real Hadoop load~\cite{OpenCloud}.
Using our proposed methodology, developers would % analysis allows to early
identify up to 78\% of tasks failures early on before the deployment of the application. Based on the performed failures analysis, our methodology could provide potential strategies to overcome these failures. \Mbarka{For instance, our solution could propose new scheduling strategies to adjust the Hadoop cluster settings (\eg{} size of the queue, allocated resources to the queues in the scheduler, etc.) and reduce the failures rate when compared to the real-execution simulation results.}
% of (up to 78\%) when compared to the real-execution simulation traces.
%\Foutse{really?how? I thought the result would help practitioner better assign resources to mitigate the failures...to propose novel scheduling strategies we should propose an extension of existing schedulers like we did for  ATLAS, but is it what we do in this paper?}
To the best of our knowledge, the present work is different from existing research in applying and integrating both formal methods and simulation for the analysis of Hadoop schedulers and formally analyzing the impact of the scheduling decisions on the failures rate in Hadoop. Furthermore, our proposed methodology to integrate model checking and simulation to verify Hadoop schedulers could be also applied to formally analyze other schedulers, than Hadoop, such as Spark~\cite{SparkOnline}, which has become one of the key cluster-computing framework that can be running on Hadoop.\\
%One can adapt the proposed methodology according to the architecture of Spark. \\
%\Foutse{in which aspect? be precise please....}.
\indent The rest of the paper is organized as follows:
Section~\ref{preliminaries} describes basics of Hadoop architecture, CSP language, and the tool PAT.
Section~\ref{framework} presents the proposed methodology for the formal analysis of the Hadoop scheduler.
We describe the analysis of the scheduler of the OpenCloud a Hadoop cluster in Section~\ref{casestudy}.
Section~\ref{relatedwork} summarizes the most relevant related work.
Finally, Section~\ref{conclusion} concludes the paper and highlights some future directions.

%% file: preliminaries.tex
\section{Preliminaries}
\label{preliminaries}
\vspace{-5pt}
In this section we briefly present the Hadoop architecture and some of the basic of the CSP language and the tool PAT, which will be used in the rest of the paper. This is in order to better understand the different steps of our proposed methodology. 
\vspace{-5pt}

\subsection{Hadoop Architecture}
Hadoop~\cite{SOUALHIA2017} is an open source implementation of the MapReduce programming model~\cite{WOHA2014}. MapReduce is designed to perform parallel processing of large datasets using a large number of computers. A MapReduce job is comprised of two functions: a map and reduce, and the input data~\cite{SOUALHIA2017}.
Hadoop has become the \textit{de facto} standard for processing large data
in today's cloud environment. It is a master-slave-based framework, the master node consists of a JobTracker and NameNode. The worker/slave node consists of a TaskTracker and DataNode. The Hadoop Distributed File System (HDFS) is the storage unit responsible for controlling access to data in Hadoop.
Hadoop is equipped with a centerpiece; the scheduler which distributes tasks across worker nodes according to their availability. The default scheduler in Hadoop is based on the First In First Out (FIFO) principle.
Facebook and Yahoo! have developed two new schedulers for Hadoop: Fair scheduler and Capacity scheduler, respectively~\cite{SOUALHIA2017}.

\subsection{CSP and PAT}
CSP is a formal language used to model and analyze the behavior of processes in concurrent systems. It has been practically applied in modeling several real time systems and protocols~\cite{CSP}.   
In the sequel, we present a subset of the CSP language, which will be used in this work, where \textit{P} and \textit{Q} are processes, \textit{a} is an event, \textit{c} is a channel, and \textit{e} and \textit{x} are values:\\
\texttt{\normalsize{
P , Q ::= Stop $\mid$ Skip $\mid$ a $\rightarrow$ P $\mid$ P ; Q $\mid$ P $\mid\mid$ Q $\mid$ c!e $\rightarrow$ P $\mid$ c?x $\rightarrow$ P
}} 
\begin{itemize}
\item \texttt{Stop}: indicates that a process is in the state of deadlock.
\item \texttt{Skip}: indicates a successfully terminated process.
\item \texttt{a $\rightarrow$ P}: means that an object first engages in the event \textit{a} and then behaves exactly as described by \textit{P}.
\item \texttt{P ; Q}: denotes that \textit{P} and \textit{Q} are sequentially executed.
\item \texttt{P $\mid\mid$ Q}: denotes that \textit{P} and \textit{Q} are processed in parallel. The two processes are synchronized with the same communication events.
\item \texttt{c!e $\rightarrow$ P}: indicates that a value \textit{e} was sent through a channel \textit{c} and then a process \textit{P}.
\item \texttt{c?x $\rightarrow$ P}: indicates a value was received through a channel \textit{c} and stored in a variable \textit{x} and then a process \textit{P}.
\end{itemize}
\noindent PAT~\cite{PAT} is a CSP-based tool used to simulate and verify concurrent, real-time systems, etc.~\cite{Sun2009}. It implements different model checking techniques for system analysis and properties verification in distributed systems (\eg{} deadlock-freeness, reachability, etc.). Different advanced optimizations techniques, such as partial order reduction, symmetry reduction, etc., are available in PAT to reduce the number of explored states and CPU time.

%% file: framework.tex
\section{Formal Analysis Methodology}
\label{framework}
\vspace{-5pt}
In this section, we first present a general overview of our methodology followed by a description of each step.

\vspace{-5pt}
\subsection{Methodology Overview}
\begin{figure}[ht]
\centering
\vspace{-10pt}
\includegraphics[scale=0.5]{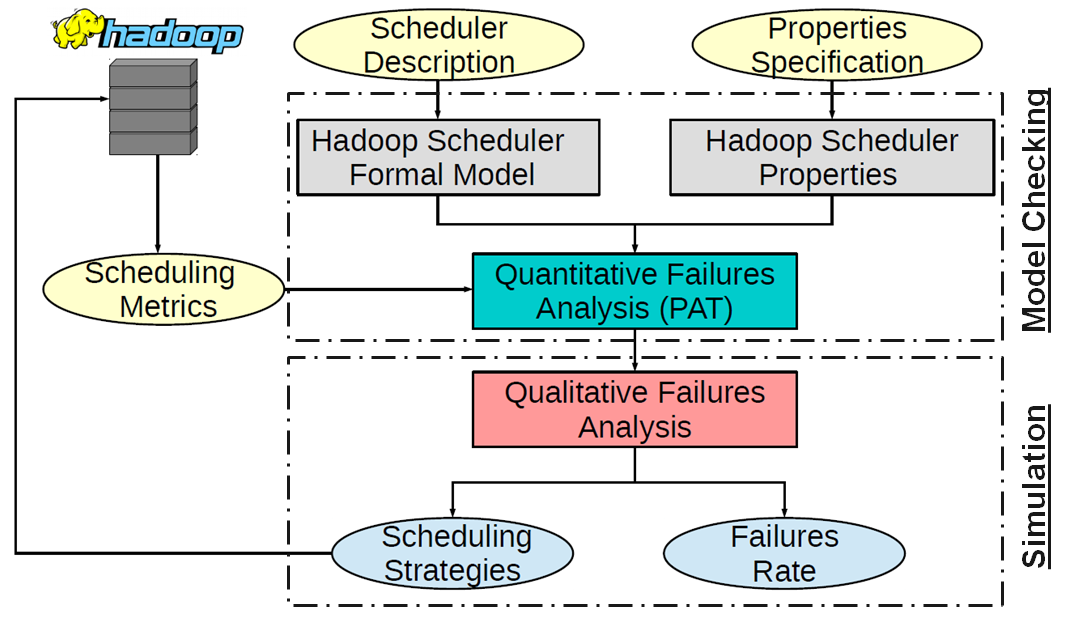}
\caption{Overview of the Formal Analysis of Hadoop Schedulers Methodology}
\vspace{-15pt}
\label{Figure:flowdiagram}
\end{figure}

Figure~\ref{Figure:flowdiagram} provides an overview of the main idea behind our formal analysis of the Hadoop scheduler when integrating model checking and simulation, to early identify failure occurrences.
The inputs of our proposed methodology are (1) the description of the Hadoop scheduler, (2) the specification of the properties to be verified, and (3) the scheduling metrics (\eg{} type of scheduler, number of nodes, workload and failure distributions, schedulability rate).
Our proposed methodology is comprised of three main steps, including (a) the formal modeling of the Hadoop scheduler and its properties in CSP and LTL, (b) the quantitative analysis of failures using model checking in PAT, and (c) the qualitative analysis of the failures using simulation of the proposed scheduling strategies and real-simulation traces. The outputs of our methodology are the rate of failures of the verified scheduler, and a set of possible scheduling strategies to overcome these failures.

\noindent We have chosen the CSP language to formally describe the scheduler as it allows to model the behavior of processes in concurrent systems. It has been successfully applied in modeling synchronous and parallel components for several real-time and distributed systems~\cite{CSP} (which is the case of Hadoop). The properties we aim to verify are written in Linear Temporal Logic (LTL).
Thereafter, we use the PAT model checker to perform the formal quantitative analysis of failures in Hadoop scheduler. PAT is based on CSP and implements various model checking techniques to analyze and simulate the behavior of several distributed systems (\eg{} transportation
management system, Web authentication protocols, wireless sensor networks, etc.)~\cite{PAT}.    
%has been widely used to simulate and verify concurrent and parallel processes in distributed systems, etc.~\cite{PAT}.} 
Furthermore, it allows to model timed and probabilistic processes in the studied systems~\cite{Sun2009}. Based on the generated results from PAT, we perform a qualitative failures analysis to determine the circumstances and specifications leading to tasks' failures in the scheduler.
The remainder of this section elaborates more on each of the steps of our methodology.

\vspace{-10pt}

\subsection{Hadoop Scheduler Formal Model: Model Checking}
\vspace{-3pt}
The first step in conducting the proposed formal analysis of the Hadoop scheduler using a model checker is to construct a formal model of the main components in Hadoop responsible for the scheduling of tasks, using the CSP language.
To do so, we start by writing a formal description of the Hadoop master node: the JobTracker and NameNode.
At the master level, we model the scheduler and the main entities responsible for task assignment and resources allocation in Hadoop.
Next, we model the TaskTracker and the DataNode including the entity responsible for the task execution at the worker nodes.
In addition, we integrate some of the important scheduling constraints in Hadoop in the model of the TaskTracker nodes including the data locality, data placement, and speculative execution of tasks~\cite{SOUALHIA2017}. Here, we selected these three constraints because of their direct impact on the scheduling strategies and the performance of executed tasks~\cite{SOUALHIA2017}. 
At this level, we should mention that the provided model of the Hadoop scheduler represents a close representation of the actual ones (\eg{} FIFO, Fair, and Capacity) because it includes the main functionalities responsible for assigning the received tasks on available nodes.
Furthermore, we checked the correspondence between the provided model and the real Hadoop scheduler by comparing the scheduling outcomes of scheduled tasks when using the formal model and the existing ones. More details about this comparison is given in Section~\ref{QFA}. 
The obtained results showed a good matching between the two models. Nevertheless, further functionalities can be added to the presented model in order to improve its results (\textit{i.e.,} recovery mechanisms, efficient resources assignment). 
%\blue{here explain correspondence between the model and actual hadoop behavior ??}
%\Foutse{why only these ones? how representative are them? can you justify a bit?}
In the following, we present a formal description of the steps to model a Hadoop cluster. Then, we present examples of implemented CSP processes\footnote{The entire CSP script is available at:\\ \url{http://hvg.ece.concordia.ca/projects/cloud/fvhs.html}} to describe the \textit{Hadoop scheduler}, \textit{TaskTracker activation} and \textit{task assignment}, where ``\textit{Cluster()}" is the main process and \textit{N} is the number of available TaskTracker nodes in the cluster: 
\vspace{4pt}

\noindent{\texttt{\footnotesize{
Cluster()~=~initialize(); NameNode$\_$activate() $\parallel$ JobTracker$\_$activate() $\parallel$~\\
\hspace*{6.5em}($\parallel$~i:\{0..(N-1)\}$@$DataNode$\_$activate(i)) $\parallel$~\\
\hspace*{6.5em}($\parallel$~i:\{0..(N-1)\}$@$TaskTracker$\_$activate(i))$\parallel$~\\
\hspace*{6.5em}Hadoop$\_$Scheduler();
}} 
\vspace{4pt}

\noindent The following process presents a formal description of the steps in \textit{``Hadoop$\_$Scheduler()"} to check scheduling constraints of map/reduce tasks. First, it checks the availability of resource slots (\textit{slotTT[i]$>$0}).
Then, it checks the type of task to be scheduled, either a map (\textit{Queue[index]~==~1}) or reduce (\textit{Queue[index]~==~2}) task. It also checks whether a task is speculatively executed or not (\eg{} map: \textit{Queue[index]~==~3} or reduce: \textit{Queue[index]~==~4}). Next, it assigns the received task to the TaskTracker node where it will be executed (\textit{signedtask?i~~$\rightarrow$~~signedtask$\_$i~~$\rightarrow$~~Task$\_$Assignment(location,type);}).\\

\noindent{\texttt{\footnotesize{
Hadoop$\_$Scheduler() = \\
\hspace*{0.5em}\{\hspace*{1.2em}if(~(slotTT[i]>0)~)~~\\
\hspace*{2.1em}\{while(~(found==0)~\&\&~(index~<~maxqueue~)~)\\
\hspace*{2.7em}\{~if~(Queue[index]~==~1)~~//it~is~a~map~task~\\
\hspace*{3.5em}\{~~schedulable~=~1;~found~=1;~location~=~index;\\
\hspace*{4.5em}  type~=~MapTask;~IDjob$\_$~task~=~IDJob[index];\}~\\
\hspace*{3em}~if((Queue[index]~==~2))~~//it~is~a~reduce~task~\\
\hspace*{3.5em}\{~if(FinishedMap[IDJob[index]]~==~Map[IDJob[index]]~~)~~\\
\hspace*{4em}\{~schedulable~=~1;~found~=1;~location~=~index;\\
\hspace*{4.5em}~type~=~ReduceTask;~IDjob$\_$~task~=~IDJob[index];\}~~~~\}\\
\hspace*{3em} if(Queue[index]~==~3)~~//it~is~a~speculated~map~task~\\
\hspace*{3.5em} \{ schedulable~=~1;~found~=1;~location~=~index;~type~=~MapTask;\\
\hspace*{4.5em}~IDjob$\_$~task~=~IDJob[index];~SpeculateTask[location]~=~1;\}~\\
\hspace*{3em} if(Queue[index]~==~4)~~//it~is~a~speculated~reduce~task~\\
\hspace*{3.5em}\{ schedulable~=~1;~found~=1;~location~=~index;~type~=~ReduceTask;\\
\hspace*{4.5em}~IDjob$\_$~task~=~IDJob[index];~SpeculateTask[location]~=~1;\}~\\
\hspace*{4.5em}...\\
\hspace*{2.8em}\}~\\
\hspace*{2.2em}\}\\
\hspace*{0.7em}\hspace*{1em}\}~$\rightarrow$~~signedtask?i$\rightarrow$~~signedtask$\_$i$\rightarrow$Task$\_$Assignment(location,~type);~
}} \\

\noindent \textit{``TaskTracker$\_$activate(i)"} presents our proposed process to activate the TaskTracker, after checking that the JobTracker was already activated (\textit{JobTracker~==~ON}) and this TaskTracker was not already activated (\textit{TaskTracker[i]~==~OFF}). Next, it activates this Tasktracker (\textit{TaskTracker[i]~==~ON}) and the number of slots specified to this TaskTracker (\textit{slotsnb}). These activated slots are ready to execute tasks.
\\

\noindent{\texttt{\footnotesize{
TaskTracker$\_$activate(i)~=~activate$\_$jt$\_$success $~\rightarrow$~ifa(TaskTracker[i]~==~~\\
\hspace*{11em} OFF \&\& JobTracker~==~ON) \{activate$\_$tt.i\{\\
\hspace*{11.5em}TaskTracker[i]~=~ON; trackercount++;\}\\
\hspace*{11.5em}~$\rightarrow$~atomic\{activate$\_$tt$\_$success.i $\rightarrow$\\
\hspace*{11.5em}($\parallel$~j:\{1..(slotsnb)\}$@$TaskTracker$\_$sendready(i))\}\};
}} \\

%\noindent{\texttt{\footnotesize{
%TaskTracker$\_$sendready(i)~=~[DataNode[i]~==~ON~\&\& TaskTracker[i]~==~ON]\\
%\hspace*{12em}sendReady!i$\rightarrow$sendReady$\_$~success.i$\rightarrow$ signedtask!i\\
%\hspace*{12em}$\rightarrow$~received$\_$task.i~$\rightarrow$~TaskTracker$\_$execute(i));
%}} \\

\noindent The following \textit{``TaskTracker$\_$execute(i)"} process presents an example of executing a task after checking its locality in Hadoop. For instance, it checks the availability of the slot assigned to a given task by the scheduler (if slot is free then \textit{task$\_$running[nbTT][k]} is equal to 0, where \textit{nbTT} is the ID of the TaskTracker and \textit{k} is the ID of the assigned slot). Next, it checks the locality of the task by checking whether the node where to execute the task (\textit{selectedTT}) is the same node where its data is located (\textit{Data-LocalTT[idtask]}).
\\
%\newpage
\vspace{-3pt}
\noindent{\texttt{\footnotesize{
TaskTracker$\_$execute(i) =\\
\hspace*{0.3em}\{\hspace*{0.3em}var~nbTT~=i;~var~found~=~0;~var~k~=~0;~\\
\hspace*{1em}while((k<slotsnb)~\&\&~(found~==~0)~)\{~\\
\hspace*{1.5em}if(task$\_$~at$\_$~tasktracker[nbTT][k]==1~\&\&~task$\_$running[nbTT][k]==~0)~\\
\hspace*{2em}\{selectedslot~=k;~found~=~1;~\}~\\
\hspace*{1.5em}k++;~\}~...\\
%\hspace*{1.5em}...\\
\hspace*{1.5em}if(Data-LocalTT[idtask]~==~selectedTT)~//check~locality~of~the~task~\\
\hspace*{2em}\{locality~=~locality~+~1;~~Locality[idtask]~=~1;\}~\\
\hspace*{2.5em}else~\{nonlocality~=~nonlocality~+~1;~~Locality[idtask]~=~0;\hspace*{1.5em}\}~\\
\hspace*{1em}...\}~\\
\hspace*{0.3em}~\}~$\rightarrow$~if(pos==~-1)~\{TaskTracker$\_$~execute(i)\}\\
\hspace*{5.5em}~else~\{execute(i,selectedslot)\};~\\
}}
\vspace{-15pt}
\subsection{Hadoop Scheduler Properties: Model Checking}
The three selected properties we aim to verify in our work are the schedulability, fairness and resources-deadlock freeness. We select these properties because they represent some of the main critical properties affecting the execution of tasks in real-time systems (\eg{} task outcome, delays, resources utilization)~\cite{Cheng2002}. The three properties can be described as follows: the \textit{schedulability} checks whether a task is scheduled and satisfies its specified deadline when scheduled according to a scheduling algorithm.
The \textit{fairness} checks whether the submitted tasks are served (\eg{} receiving resources slots that ensure their processing) and there are no tasks that will never be served or will wait in the queue more than expected.
% for long time \Foutse{how long?}.
The \textit{resources-deadlock} checks whether two tasks are competing for the same resource or each is waiting for the other to be finished.

To better illustrate the properties to be verified, we explain as an example the schedulability property and its corresponding states in our approach.
For the schedulability, a task can go from state: \textit{submitted} to \textit{scheduled} to \textit{processed} then to \textit{finished-within-deadline} or \textit{finished-after-deadline} or \textit{failed}.
Let \textit{X} be the total number of scheduled tasks and \textit{Y} be the number of tasks finished within their expected deadlines. The schedulability rate can be defined as the ratio of \textit{X} over \textit{Y}. The property \textit{``schedulabilityrate $ > $ 80"} means checking whether the scheduler can have a total of 80\% of tasks finished within their deadlines.

To verify above properties in PAT, we need to provide their descriptions in LTL. For example, the following LTL formulas check the schedulability and resource-deadlock freeness of given tasks. The first example checks whether a given task eventually goes from the state submitted to the state finished within the deadline. The second example checks whether a given task should not go to a state of waiting-resources. Here, $\diamondsuit$, $\models$, -$>$, and $ \lnot $ represent \textit{eventually}, \textit{satisfy}, \textit{imply}, and \textit{not}, respectively, in the LTL logic. \\

\noindent{\texttt{\footnotesize{
\hspace*{-0.6em}$\#$~assert $\diamondsuit$ (task~$\models$~(submitted -$>$ finished-within-deadline) );\\
%\hspace*{13em}~finished-within-deadline) );\\
$\#$~assert $ \lnot $ (task~$\models$~(submitted -$>$ waiting-resources) );\\ 
%\hspace*{13em}~waiting-resources) );\\
}} 
\vspace{-15pt}
\subsection{Quantitative Failures Analysis using Model Checking}
\Mbarka{Provided the CSP model of the Hadoop scheduler and the properties to be verified, we perform the formal analysis of the scheduler performance using the PAT model checker while simulating different Hadoop workload scenarios.
Here, we can vary the property requirements to assess the performance of the scheduler under different rates and evaluate their impact on the failures rates of the cluster and the simulated scenarios.}
For example, we can define \textit{``goal0"} to check whether all the submitted tasks (\textit{``workload"}) are successfully scheduled. Using PAT, we can verify if the modeled cluster, \textit{``cluster1"}, can reach this goal or not. Another example could be to check if \textit{``cluster1"} meets \textit{``goal0"} with a \textit{``schedulabilityrate"} of 80\%. The following examples present some of the properties that can be verified using our approach.\\
\vspace{-23pt}
% \Foutse{we should explain these examples...and provide a link to a technical report where there is an extensive description of the properties!}:
\begin{table}[ht]
%\centering
\footnotesize
\label{my-label}
\texttt{{
\begin{tabular}{l}
$\#$define~goal0~completedscheduled~~==~workload~\&\&~workload~>0;\\
$\#$assert~cluster1~reaches~goal0;\\
%$\#$assert~cluster~reaches~goal0~with~min(schedulabilityrate);\\
%$\#$assert~cluster~reaches~goal0~with~max(schedulabilityrate);\\
$\#$assert~cluster1~reaches~goal0~~\&\&~schedulabilityrate~>80;\\
$\#$define~goal1~fairnessrate~==50;\\
$\#$assert~cluster1~reaches~goal0~~\&\&~goal1;\\
$\#$define~goal2~resourcedeadlockrate~==50;\\
$\#$assert~cluster1~reaches~goal0~~\&\&~goal1~~\&\&~goal2;
\end{tabular}
}}
\end{table}
\vspace{-20pt}

\subsection{Qualitative Failures Analysis using Simulation}
\Mbarka{The last step in our proposed methodology is to use the traces simulated and generated by the PAT model checker to extract information about the applied scheduling strategies and explore their impact on tasks' failures using simulation.} For instance, we can investigate the states where the scheduler does (not) meet a given property and map these states to the obtained scheduler performance and to the input cluster settings. This step allows us to find a possible correlation between the cluster settings, the applied scheduling strategies, and the failures rate. Next, we use these correlations to suggest scheduling strategies to overcome these failures by either (1) recommending to the Hadoop developers to change the scheduling decisions (\eg{} delay long tasks, wait for a local task execution) or (2) to customers to change and adjust their cluster settings (\eg{} number of nodes, number of allowed speculative executions). In next section, we propose a case study to evaluate and simulate the suggested scheduling strategies on a Hadoop environment and measure their impact on the failure rates and the Hadoop cluster performance. 
\Mbarka{Generally, our solution could propose new scheduling strategies to adjust the Hadoop cluster settings and hence reduce failures rate when compared to the real-execution simulation results.}

%% file: casestudy.tex
\section{Case Study: Formal Analysis of OpenCloud Scheduler}
\label{casestudy}
%\vspace{-10pt}
\Mbarka{In this section, we illustrate the practical usability and benefits of our work by formally analyzing the scheduler of OpenCloud, a Hadoop-based cluster~\cite{OpenCloud} that simulates the Hadoop load.}
\vspace{-5pt}
\subsection{Case Study Description}
%\vspace{-10pt}
To apply our formal approach on a case study, we investigated existing Hadoop case studies in the open literature.
Overall, we found three main public case studies including: Google~\cite{Google}, Facebook~\cite{Facebook}, and OpenCloud~\cite{OpenCloud} traces. %, that we checked to identify whether they contain enough inputs data of our methodology.
We found out that the Google traces do not provide data about the cluster settings, which is an important factor affecting the analysis results in our approach. For the Facebook traces, we noticed that they do not provide data about the cluster settings, capacity of nodes, failures rate, etc., which are essential for our approach.
Whereas, we found that the OpenCloud traces provide the required inputs of our verification approach (\eg{} \# nodes, capacity of nodes, workload). Therefore, we choose to formally analyze the scheduler of this cluster because it provides public traces of real-executed Hadoop workload for more than 20 months.
OpenCloud~\cite{OpenCloud} is an open research cluster managed by Carnegie Mellon University. It is used by several groups in different areas such as machine learning, astrophysics, biology, cloud computing, etc. \\
%\Foutse{what is the type of this scheduler? FIFO? capacity?...}
%Indeed, we use these traces to apply and validate our proposed methodology, comparing \red{the obtained verification results} with the ones of the real execution of the workload in terms of performance and failures rate. Our aim here is to check whether our proposed methodology could allow for an early identification of the failures that occurred in the field (i.e., failures that are reported in execution traces).\\
\indent Based on the modeled system of the OpenCloud's scheduler and the specified properties, we start the verification by parsing the trace files to extract the input information needed in our methodology.
%\Foutse{i though you parsed the trace files to extract the information needed by your framework?}.
Specifically, we use the description of the workload included in the first month traces, and the first six months traces together. This allows us to evaluate the scalability of our methodology, in terms of number of visited states and execution time, using different traces. % and its scalability.
Although these traces provide the required inputs for our methodology, we did not find any information describing the type of scheduler used in the cluster. Since the type of the scheduler is an important factor that impacts the performance of the cluster, we evaluate the performance of the modeled cluster for the three existing schedulers of Hadoop (FIFO, Fair and Capacity).
%\Foutse{how exactly? explain your assumptions...and dont forget to mention this in threats to validity!}.
%%%% ==> I guess that adding a section about threats will not be appreciated by NFM reviewers ;-)%%%
\Mbarka{We vary the property requirements to assess the performance of the used scheduler under different rates and evaluate their impact on the failures rate in the cluster while executing the same Hadoop workload.}
For the search engines in PAT, we use the ``First Witness Trace using Depth First Search" in order to perform an analysis without reduction, and the ``First Witness Trace with Zone Abstraction" for the analysis with symmetry reduction. The testbed is a workstation with Intel i7-6700HQ (2.60GHz*8) CPU and 16 GB of RAM. 
\vspace{-5pt}
\subsection{Verification and Evaluation}
\vspace{-5pt}
In this section, we present the obtained scalability results of our methodology along with the results of the formal quantitative and qualitative analyses of failures in Hadoop schedulers. 
\vspace{-10pt}
%\Foutse{have at least a sentence before diving into sub-sections...}
\subsubsection{Properties Verification and Scalability Analysis}
\label{perfanalysis}
Tables~\ref{Results-File1} and~\ref{Results-File3} summarize the verification results of the first month trace, and the first six  months traces together, respectively. The first trace provides information about 1,772,144 scheduled tasks, whereas the six files describing the executed workload for six months contain information about 4,006,512 scheduled tasks.
In the sequel, we discuss the results of the performed analysis. \\
\indent The results presented in Table~\ref{Results-File1} show that only the Fair scheduler satisfies the schedulability property (for the two given rates), meaning that up to 80\% are scheduled and executed within their expected deadlines. The Capacity scheduler does not meet the schedulability rate of 80\%.
However, the FIFO does not satisfy the schedulability properties for the two input rates, meaning that more than 50\% of scheduled tasks are exceeding their expected deadlines. Hence, these tasks are using their assigned resources more than expected, which can affect the overall performance of the cluster.  \\
\indent For the fairness property, only the Fair scheduler satisfies the property of fairness with a rate of 50\% and 80\%, meaning that at most 80\% of tasks get served and executed on time. However, both the FIFO and the Capacity schedulers violate the fairness property for the two given values. Therefore, we can claim that more than 80\% of the submitted tasks are waiting longer than expected in the queue before getting executed. This may lead to a problem of task starvation.
 %, especially when tasks are waiting in the queue for long time.

\input{table1}
\indent For the resources-deadlock, we can report that the Capacity scheduler is characterized by more tasks that experience a resources-deadlock compared to the FIFO and the Fair schedulers. This is because it satisfies the property of resources-deadlock for the two given rates (\eg{} 10\% and 30\%). This can be explained by the fact that the Capacity scheduler suffers from the problem of miscalculation of resources (headroom calculation).
The FIFO scheduler shows that only 10\% of the scheduled tasks experience the problem of resources-deadlock.
The Fair scheduler does not satisfy the resources-deadlock property for two given rates (10\% and 30\%) for the first trace, which means that less than 10\% of tasks may experience an issue of resources-deadlock.
Hence, we can conclude that the Fair scheduler generates less scheduling decisions leading to resources-deadlock situations for the OpenCloud cluster.\\
\indent Overall, our proposed methodology is able to verify the given three properties for the three existing Hadoop schedulers, while exploring on average 11086 K states in 3724 seconds without symmetry reduction, and 742 K states in 1619 seconds with symmetry reduction, as shown in Table~\ref{Results-File1}.
%\input{table2}

%When applied to the second traces of the OpenCloud cluster, our proposed methodology could formally analyze the three properties for the three schedulers using raw data of 476,034 tasks.
%We observe that the FIFO, Fair and Capacity schedulers verify achieve a schedulability rate of 30\% and a fairness rate of 20\%, whereas they all do not meet a rate of 90\% of schedulability and fairness, as shown in Table~\ref{Results-File2}.
%\red{For the resources-deadlock, we notice that at least 35\% of tasks can experience resource-deadlock when scheduled with the Capacity scheduler.} %is characterized by 35\% of scheduled tasks that experienced resource-deadlock situation,
%%, while it does not meet a resource-deadlock \Foutse{what does this mean???? meeting resource-deadlock??} of 50\%.\\
%In a comparison with the first trace, our approach explores 2856 K states in (up to) 2186 seconds without symmetry reduction, and 581 K states in (up to) 1246 seconds when applying symmetry reduction. This is expected due to the huge difference in the size between the two input traces.\\
%Overall, the formal analysis of the second trace using our framework explored less states

\indent In order to evaluate the scalability of our methodology, we perform the formal analysis of a cumulative workload. This is done by incrementally adding the Hadoop workload of each month to the previous trace(s) to be analyzed. We start the evaluation by analyzing the trace of the first month, and add the traces up to the trace where the tool cannot perform the analysis. This is in order to check the bounds of the analysis performed in terms of explored states.
The obtained results, presented in Table~\ref{Results-File3}, show that our approach could formally analyze the first six traces (combined) describing the workload (about 4,006,512 tasks) for the first six months. It could analyze the first five traces together with and without symmetry reduction, however, it could analyze the six traces together
only with symmetry reduction by exploring on average 17,692 K states in 4346 seconds.
\input{table3}
%\vspace{-5pt}
\subsubsection{Quantitative Failures Analysis}
\label{QFA}
We use the traces generated by the PAT model checker during the verification of the three properties described in Section~\ref{perfanalysis}, to explore states where the scheduler did not satisfy a given property value.
This step is important to map these states to the failed tasks, if found, and examine the relationship between the verified property and the task failure. We apply this step on the first trace file because it contains an important number of scheduled tasks (i.e., 1,772,144 tasks).
To do so, we first identify the tasks that did not satisfy a given property, for example schedulability = 80\% or resources-deadlock
= 10\%. Then, we check whether these tasks were failed when they were executed in the real cluster.
Next, we classify the observations into four main categories: \textit{True Positive (TP)}, \textit{True Negative (TN)}, \textit{False Positive (FP)}, and  \textit{False Negative (FN)}.
\textit{TP} represents the successfully executed tasks, based on the simulation traces, that are identified as finished tasks using our approach.
%\Foutse{you never mentioned prediction before....please use consistent term...in your case you need a term adequate for verification studies!}
Likewise, \textit{TN} represents the failed tasks, based on the simulation traces, that are identified as failed tasks.
While \textit{FP} denotes the amount of identified finished tasks that failed during the real simulation. Finally, \textit{FN} denotes the number of identified failed tasks that are finished in reality.
These four metrics are calculated by comparing the status of the tasks in the generated traces using our methodology, specifically from PAT, and the simulation traces across the total number of scheduled tasks in the input trace (1,772,144 tasks). \\
\indent Overall, we observe that our formal analysis approach could identify up to 61.49\% of the finished tasks (\textit{TP}, Fair, schedulability = 50\%), and up to 4.62\% of the failed tasks (\textit{TN}, Fair, schedulability = 80\%) without symmetry reduction. Here, we notice that the \textit{TN} values are small compared to the \textit{TP} ones. This can be explained by the fact that the first trace contains more than 94\% of successful tasks. For this reason, we computed another metric that we call the \textit{Detected Failures (DF)}.
The \textit{DF} is calculated by mapping the \textit{TN} rate over the total number of failed tasks in the input trace.
\Mbarka{Our aim here is to quantify the amount of detected failures using our formal verification approach when compared to the given OpenCloud traces (obtained from real-execution simulation). This is to answer the question of how many failures could be identified by our formal verification approach before the application is deployed in the field?}\\ %our formal approach early identify failures when compared to the real-execution traces.
\indent At this level, we can claim that our approach is able to catch between 42\% and 78.57\% of the total failed tasks without symmetry reduction. While it can catch between 42\% and 78.91\% of the failures when using the symmetry reduction.
On the other hand, we notice that the \textit{FN} is in the order of 40\%, which means that more than 40\% of tasks are finished, in the simulation traces, but our methodology indicates their failures. This can be explained by the internal mechanisms implemented in Hadoop internally to recover these tasks in case of a failure. For example, the mechanism of pausing and resuming running tasks allows higher priority tasks to be executed without killing the lower priority ones~\cite{RAFT2011}. Furthermore, we notice that the false positive rate varies from 1.26\% and 3.41\% and indicates failed tasks that are identified as finished using our methodology. A possible explanation for these false positive results can be the lack of some real-world constraints that affect the execution of tasks (\eg{} network congestion, lack of resources). In addition, failed tasks identified as finished can generate more false positive results due to the tight dependency between the map and the reduce tasks (the reduce tasks will be launched when all the map tasks are successfully executed). 
%\vspace{-30pt}
%\Foutse{can you give an example of such recovery mechanisms?}\Foutse{say that you will investigate ways to include these recovery mechanisms in your methodology in the future!} \\
\input{table6}
%\vspace{-10pt}
%\vspace{-10pt}

\indent Next, we check the states of the failed tasks and analyze the factors that may cause the failure of these tasks.
We find that the scheduler of OpenCloud experiences several failures, up to 32\%, because of long delays that exceed the maximum timeout for a task to be finished (property ``mapred.task.timeout": defining the maximum timeout for a task to be finished).
%These tasks do not meet the schedulability requirement.
We carefully checked the sates of these tasks and found that these delays are mainly caused by data locality constraints~\cite{SOUALHIA2017} and that they can reach 10 minutes (for small and medium tasks). Moreover, we found out that about 40\% of these straggling tasks cause the failure of the job to which they belong, resulting in a waste of resources.
Moreover, we noticed that many failures are cascaded from one job to another, especially in the long Hadoop chains. A Hadoop chain is a connection between more than one Hadoop job to execute a specific workload. This may result in a degradation in performance and many failures. We can claim that the Hadoop scheduler lacks mechanisms to share information about these failures between its components to avoid their occurrence.
On the other hand, we found out that 26\% of tasks failed because they exceeded the maximum number of allowed speculative execution (\eg{} property ``mapred.map.tasks.speculative.execution": defining the maximum number of allowed speculative execution, they have likely more chances to fail).
When checking the tasks waiting for a long time in the queue before being served, we observed that this is due to the fact that the long tasks are scheduled first and they occupy their assigned resources for a long time (\eg{} a large input file to be processed, a job composed of more than 1000 tasks). Consequently, we can conclude that knowing these factors and issues, one can adapt the scheduler system of the Hadoop framework to overcome these failures and improve its performance.\\

% \Foutse{how concretely? elaborate...and possibly give some examples!}.
\vspace{-15pt}
\subsubsection{Qualitative Failures Analysis}
\begin{figure}[ht]
\centering
\includegraphics[scale=0.8]{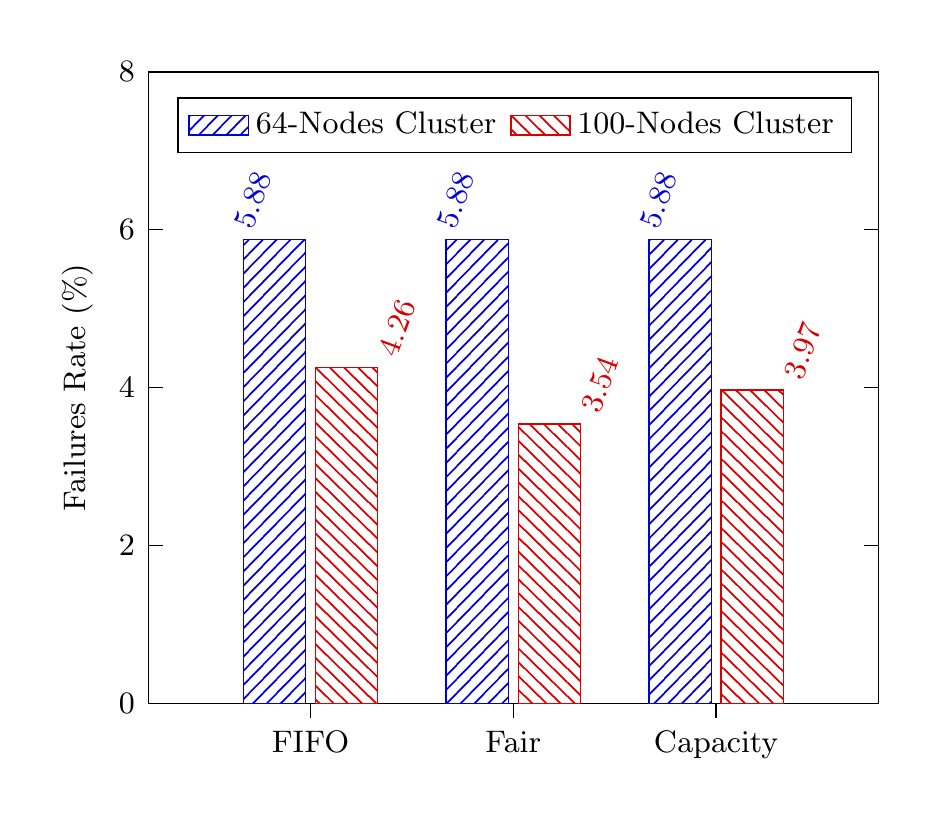}
\vspace{-10pt}
\caption{Impact of Resources Adding on Failures Rate}
%Trace for the First Month (1,772,144 Tasks)}
\label{Figure:Results}
\end{figure}

\Mbarka{To show the benefits of our formal analysis approach, we propose to integrate and simulate some guidelines or strategies to adapt the scheduling decisions of the created Hadoop cluster and evaluate their impact on the failures rate.} This is based on the generated traces and the performed failures analysis to check whether our work can help early identify the occurrence of failures and then propose appropriate mechanisms to overcome them.
For instance, one possible strategy could be to change the cluster settings by adding more resources on its nodes or adding the number of nodes on it. This could solve for example the fairness and resources-deadlock issues. Another scheduling strategy could be to change the value of timeout of scheduled task, which represents the number of milliseconds before terminating a task. Another strategy could be to adjust the type of the scheduler used in the cluster (\eg{} FIFO, Fair, Capacity, etc.).
%\Foutse{this can have downsides since tasks that will fail anyway will consume more resources that they should...so we should discuss these aspects!...}.
\Mbarka{In this paper, we evaluate the impact of the strategy to change the cluster settings by adding more resources on the OpenCloud cluster where we change the number of nodes from 64 to 100 and simulate the same Hadoop workload}. Figure~\ref{Figure:Results} gives an overview of the impact of adding more resources in the cluster for the three schedulers considering a failures' rate of 5.88\%; the identified failure rate from the first trace file.
% \Foutse{what do you mean by "such that the initial failures rate is"?}.
Overall, we noticed that adding more nodes in the cluster could reduce the failure rates by up to 2.34\% (Fair scheduler), which means a reduction rate of 39.79\%.
%\Foutse{this is the reduction rate or the new failure rate after addition of resources?}  
This was expected since when we analyzed the traces we found out that several tasks are straggling for more than 10 minutes, waiting for other resources to be released. \\ 
\Mbarka{\indent Given the obtained findings, we can claim that our solution could identify new scheduling strategies to adjust the Hadoop cluster to reduce failures rates by combining simulation and model checking techniques and formally verifying the functional scheduling properties.}

%\vspace{-30pt}
%\vspace{-20pt}

%% file: table1.tex
%\vspace{-20pt}
\begin{table}[ht]
\centering \scriptsize
\caption{Verification Results: Trace for the First Month (1,772,144 Tasks)}
\label{Results-File1}
%\resizebox{\textwidth}{!}{%
\begin{tabular}{|c|c|c|c|c|c|c|c|}
\hline
\multirow{2}{*}{\textbf{Property}}                                                               & \multirow{2}{*}{\textbf{Scheduler}} & \multicolumn{3}{c|}{\textbf{Results without SR}}       & \multicolumn{3}{c|}{\textbf{Results with SR}}          \\ \cline{3-8} 
                                                                                                 &                                     & \textbf{Valid?} & \textbf{\#States} & \textbf{Time(s)} & \textbf{Valid?} & \textbf{\#States} & \textbf{Time(s)} \\ \hline\hline
\multirow{3}{*}{\textbf{\begin{tabular}[c]{@{}c@{}}Schedulability\\ = 50\%\end{tabular}}}        & \textbf{FIFO}                       & No              & 11086 K           & 3869             & No              & 742 K             & 1648             \\ \cline{2-8} 
                                                                                                 & \textbf{Fair}                       & Yes             & 11086 K           & 3524             & Yes             & 742 K             & 1597             \\ \cline{2-8} 
                                                                                                 & \textbf{Capacity}                   & Yes             & 11086 K           & 3601             & Yes             & 742 K             & 1604             \\ \hline
\multirow{3}{*}{\textbf{\begin{tabular}[c]{@{}c@{}}Schedulability\\ = 80\%\end{tabular}}}        & \textbf{FIFO}                       & No              & 11086 K           & 3789             & No              & 742 K             & 1650             \\ \cline{2-8} 
                                                                                                 & \textbf{Fair}                       & Yes             & 11086 K           & 3676             & Yes             & 742 K             & 1614             \\ \cline{2-8} 
                                                                                                 & \textbf{Capacity}                   & No              & 11086 K           & 3721             & No              & 742 K             & 1602             \\ \hline\hline
\multirow{3}{*}{\textbf{\begin{tabular}[c]{@{}c@{}}Fairness\\ = 50\%\end{tabular}}}              & \textbf{FIFO}                       & No              & 11086 K           & 3714             & No              & 742 K             & 1594             \\ \cline{2-8} 
                                                                                                 & \textbf{Fair}                       & Yes             & 11086 K           & 3759             & Yes             & 742 K             & 1615             \\ \cline{2-8} 
                                                                                                 & \textbf{Capacity}                   & No              & 11086 K           & 3736             & No              & 742 K             & 1612             \\ \hline
\multirow{3}{*}{\textbf{\begin{tabular}[c]{@{}c@{}}Fairness\\ = 80\%\end{tabular}}}              & \textbf{FIFO}                       & No              & 11086 K           & 3855             & No              & 742 K             & 1675             \\ \cline{2-8} 
                                                                                                 & \textbf{Fair}                       & Yes             & 11086 K           & 3795             & Yes             & 742 K             & 1642             \\ \cline{2-8} 
                                                                                                 & \textbf{Capacity}                   & No              & 11086 K           & 3761             & No              & 742 K             & 1619             \\ \hline\hline
\multirow{3}{*}{\textbf{\begin{tabular}[c]{@{}c@{}}Resources-\\ Deadlock\\ = 10\%\end{tabular}}} & \textbf{FIFO}                       & Yes             & 11086 K           & 3615             & Yes             & 742 K             & 1602             \\ \cline{2-8} 
                                                                                                 & \textbf{Fair}                       & No              & 11086 K           & 3698             & No              & 742 K             & 1610             \\ \cline{2-8} 
                                                                                                 & \textbf{Capacity}                   & Yes             & 11086 K           & 3704             & Yes             & 742 K             & 1632             \\ \hline
\multirow{3}{*}{\textbf{\begin{tabular}[c]{@{}c@{}}Resources-\\ Deadlock\\ = 30\%\end{tabular}}} & \textbf{FIFO}                       & No              & 11086 K           & 3742             & No              & 742 K             & 1596             \\ \cline{2-8} 
                                                                                                 & \textbf{Fair}                       & No              & 11086 K           & 3733             & No              & 742 K             & 1618             \\ \cline{2-8} 
                                                                                                 & \textbf{Capacity}                   & Yes             & 11086 K           & 3752             & Yes             & 742 K             & 1623             \\ \hline
\end{tabular}
%}
\begin{tablenotes}
  \item SR = Symmetry Reduction
    %\item[a] Another footnote.
  \end{tablenotes}
\end{table}
%\vspace{-10pt}

%% file: table3.tex
%\vspace{-20pt}
\begin{table}[ht]
\centering \scriptsize
\caption{Verification Results: Trace for the 1-6 Months (4,006,512 Tasks)}
\label{Results-File3}
%\resizebox{\textwidth}{!}{%
\begin{tabular}{|c|c|c|c|c|c|c|c|}
\hline
\multirow{2}{*}{\textbf{Property}}                                                               & \multirow{2}{*}{\textbf{Scheduler}} & \multicolumn{3}{c|}{\textbf{Results without SR}}       & \multicolumn{3}{c|}{\textbf{Results with SR}}          \\ \cline{3-8} 
                                                                                                 &                                     & \textbf{Valid?} & \textbf{\#States} & \textbf{Time(s)} & \textbf{Valid?} & \textbf{\#States} & \textbf{Time(s)} \\ \hline\hline
\multirow{3}{*}{\textbf{\begin{tabular}[c]{@{}c@{}}Schedulability\\ = 50\%\end{tabular}}}        & \textbf{FIFO}                       & **              & **                & **               & Yes             & 17692K            & 4350             \\ \cline{2-8} 
                                                                                                 & \textbf{Fair}                       & **              & **                & **               & Yes             & 17692K            & 4362             \\ \cline{2-8} 
                                                                                                 & \textbf{Capacity}                   & **              & **                & **               & Yes             & 17692K            & 4359             \\ \hline
\multirow{3}{*}{\textbf{\begin{tabular}[c]{@{}c@{}}Schedulability\\ = 90\%\end{tabular}}}        & \textbf{FIFO}                       & **              & **                & **               & No              & 17692K            & 4346             \\ \cline{2-8} 
                                                                                                 & \textbf{Fair}                       & **              & **                & **               & No              & 17692K            & 4341             \\ \cline{2-8} 
                                                                                                 & \textbf{Capacity}                   & **              & **                & **               & No              & 17692K            & 4367             \\ \hline\hline
\multirow{3}{*}{\textbf{\begin{tabular}[c]{@{}c@{}}Fairness\\ = 30\%\end{tabular}}}              & \textbf{FIFO}                       & **              & **                & **               & Yes             & 17692K            & 4377             \\ \cline{2-8} 
                                                                                                 & \textbf{Fair}                       & **              & **                & **               & Yes             & 17692K            & 4312             \\ \cline{2-8} 
                                                                                                 & \textbf{Capacity}                   & **              & **                & **               & Yes             & 17692K            & 4335             \\ \hline
\multirow{3}{*}{\textbf{\begin{tabular}[c]{@{}c@{}}Fairness\\ = 90\%\end{tabular}}}              & \textbf{FIFO}                       & **              & **                & **               & No              & 17692K            & 4352             \\ \cline{2-8} 
                                                                                                 & \textbf{Fair}                       & **              & **                & **               & No              & 17692K            & 4328             \\ \cline{2-8} 
                                                                                                 & \textbf{Capacity}                   & **              & **                & **               & No              & 17692K            & 4369             \\ \hline\hline
\multirow{3}{*}{\textbf{\begin{tabular}[c]{@{}c@{}}Resources-\\ Deadlock\\ = 10\%\end{tabular}}} & \textbf{FIFO}                       & **              & **                & **               & Yes             & 17692K            & 4322             \\ \cline{2-8} 
                                                                                                 & \textbf{Fair}                       & **              & **                & **               & No              & 17692K            & 4360             \\ \cline{2-8} 
                                                                                                 & \textbf{Capacity}                   & **              & **                & **               & Yes             & 17692K            & 4354             \\ \hline
\multirow{3}{*}{\textbf{\begin{tabular}[c]{@{}c@{}}Resources-\\ Deadlock\\ = 50\%\end{tabular}}} & \textbf{FIFO}                       & **              & **                & **               & No              & 17692K            & 4338             \\ \cline{2-8} 
                                                                                                 & \textbf{Fair}                       & **              & **                & **               & No              & 17692K            & 4342             \\ \cline{2-8} 
                                                                                                 & \textbf{Capacity}                   & **              & **                & **               & No              & 17692K            & 4328             \\ \hline
\end{tabular}
%}
\begin{tablenotes}
  \item SR = Symmetry Reduction, ** = Not Available
    %\item[a] Another footnote.
  \end{tablenotes}
  \vspace{-20pt}
\end{table}
%\vspace{-10pt}

%% file: table6.tex
% \usepackage{multirow}
\begin{table}[ht]
\centering \scriptsize
\caption{Coverage Results(\%): Trace for the First Month (1,772,144 Tasks)}
\label{Results-Coverage}
\resizebox{\textwidth}{!}{%
\begin{tabular}{|c|c|> {\centering\arraybackslash}p{0.8cm}|> {\centering\arraybackslash}p{0.8cm}|> {\centering\arraybackslash}p{0.8cm}|> {\centering\arraybackslash}p{0.8cm}|> {\centering\arraybackslash\columncolor[gray]{0.9}}p{0.8cm}||> {\centering\arraybackslash}p{0.8cm}|> {\centering\arraybackslash}p{0.8cm}|> {\centering\arraybackslash}p{0.8cm}|> {\centering\arraybackslash}p{0.8cm}|> {\centering\arraybackslash\columncolor[gray]{0.9}}p{0.8cm}|}
\hline
\multirow{2}{*}{\textbf{Property}}                                                               & \multirow{2}{*}{\textbf{Scheduler}} & \multicolumn{5}{c||}{\textbf{Results without SR}}      & \multicolumn{5}{c|}{\textbf{Results with SR}}         \\ \cline{3-12} 
                                                                                                 &                                     & \textbf{TP} & \textbf{TN} & \textbf{FP} & \textbf{FN} & \textbf{DF} &\textbf{TP} & \textbf{TN} & \textbf{FP} & \textbf{FN} & \textbf{DF}\\ \hline\hline
\multirow{3}{*}{\textbf{\begin{tabular}[c]{@{}c@{}}Schedulability\\ = 50\%\end{tabular}}}        & \textbf{FIFO}                   & 56.03       & 3.2         & 2.68        & 38.09  &  54.76   & 47.29       & 2.47        & 3.41        & 46.83   &  42.00  \\ \cline{2-12} 
                                                                                                 & \textbf{Fair}                      & 61.49       & 4.53        & 1.35        & 32.63   &   77.04 & 55.38       & 3.82        & 2.06        & 38.74  &    64.96 \\ \cline{2-12} 
                                                                                                 & \textbf{Capacity}                 & 49.16       & 2.47        & 3.41        & 44.96   &  42.00  & 46.09       & 2.85        & 3.03        & 48.03  &    48.46 \\ \hline
\multirow{3}{*}{\textbf{\begin{tabular}[c]{@{}c@{}}Schedulability\\ = 80\%\end{tabular}}}        & \textbf{FIFO}                      & 50.14       & 3.46        & 2.42        & 43.98  &  58.84   & 47.98       & 2.79        & 3.09        & 46.14     &47.44  \\ \cline{2-12} 
                                                                                                 & \textbf{Fair}                    & 59.83       & 4.62        & 1.26        & 34.29    &  78.57 & 56.82       & 4.21        & 1.67        & 37.3  &     71.59 \\ \cline{2-12} 
                                                                                                 & \textbf{Capacity}                 & 43.61       & 3.05        & 2.83        & 37.73    & 51.87  & 42.18       & 3.03        & 2.85        & 51.94 &    51.53 \\ \hline\hline
\multirow{3}{*}{\textbf{\begin{tabular}[c]{@{}c@{}}Fairness\\ = 50\%\end{tabular}}}              & \textbf{FIFO}                    & 49.28       & 2.92        & 2.96        & 44.84  &  49.65   & 47.84       & 2.92        & 2.96        & 46.28     & 49.65 \\ \cline{2-12} 
                                                                                                 & \textbf{Fair}                     & 55.36       & 3.89        & 1.99        & 38.76     & 66.15 & 49.63       & 3.86        & 2.02        & 44.49  &65.64     \\ \cline{2-12} 
                                                                                                 & \textbf{Capacity}        & 47.03       & 2.47        & 3.41        & 47.09       &42.00 & 44.11       & 3.04        & 2.84        & 50.01   &51.70    \\ \hline
\multirow{3}{*}{\textbf{\begin{tabular}[c]{@{}c@{}}Fairness\\ = 80\%\end{tabular}}}              & \textbf{FIFO}                        & 44.88       & 3.76        & 2.12        & 49.29   &  63.94  & 49.77       & 3.32        & 2.56        & 44.35    & 56.46  \\ \cline{2-12} 
                                                                                                 & \textbf{Fair}                      & 57.21       & 4.07        & 1.81        & 36.91  &    69.21  & 50.14       & 4.64        & 1.24        & 43.98  &    78.91 \\ \cline{2-12} 
                                                                                                 & \textbf{Capacity}            & 43.14       & 3.01        & 2.87        & 50.98    &  51.19 & 48.29       & 3.48        & 2.4         & 45.83  &76.18     \\ \hline\hline
\multirow{3}{*}{\textbf{\begin{tabular}[c]{@{}c@{}}Resources-\\ Deadlock\\ = 10\%\end{tabular}}} & \textbf{FIFO}                       & 43.73       & 3.06        & 2.82        & 50.39    &  52.04 & 46.21       & 3.19        & 2.69        & 47.92   &  54.25  \\ \cline{2-12} 
                                                                                                 & \textbf{Fair}                     & 49.99       & 3.82        & 2.06        & 44.13    & 64.96  & 51.44       & 3.63        & 2.25        & 42.68   &   61.73 \\ \cline{2-12} 
                                                                                                 & \textbf{Capacity}                  & 46.07       & 3.19        & 2.69        & 48.05    & 54.25  & 43.77       & 2.84        & 3.04        & 50.35  &    48.29 \\ \hline
\multirow{3}{*}{\textbf{\begin{tabular}[c]{@{}c@{}}Resources-\\ Deadlock\\ = 30\%\end{tabular}}} & \textbf{FIFO}                       & 50.22       & 3.53        & 2.35        & 43.9   &  60.03   & 48.54       & 2.91        & 2.97        & 45.58    & 49.48  \\ \cline{2-12} 
                                                                                                 & \textbf{Fair}               & 52.39       & 4.17        & 1.71        & 41.73   &  70.91  & 55.26       & 3.75        & 2.13        & 38.86   & 63.77   \\ \cline{2-12} 
                                                                                                 & \textbf{Capacity}                 & 49.02       & 3.16        & 2.72        & 45.1    &  53.74  & 49.71       & 3.19        & 2.69        & 44.41   &54.25    \\ \hline
\end{tabular}
}
\begin{tablenotes}
  \item SR =  Symmetry Reduction, DF = Detected Failures
  \item TP = True Positive, TN = True Negative, FP = False Positive, FN = False Negative
  \end{tablenotes}
\vspace{-15pt}
\end{table}

%% file: relatedwork.tex
\section{Related Work}
\label{relatedwork}
In this section, we discuss the most relevant work that apply formal methods in the context of Hadoop framework.
In~\cite{Performance2015} and~\cite{Petrinets2015}, the authors analyzed the behavior of MapReduce using Stochastic Petri Nets and Timed Colored Petri Nets, respectively. They modeled the mean delay in each time transition of the scheduled tasks as formulas, and the Hadoop jobs were simulated based on the used Petri Nets. The proposed approaches could evaluate the correctness of the system and analyze the performance of MapReduce. But, they lack several constraints about the scheduling of the jobs and cannot cover larger Hadoop clusters.
Su \textit{et al.}~\cite{Su2009} used the CSP language to formalize four key components in MapReduce. Specifically, they formally modeled the master, mapper, reducer and file system while considering the basic operations in Hadoop (\eg{} task state storing, error handling, progress controlling). However, none of the properties of the Hadoop framework is verified using the formalized components.
Xie \textit{et al.}~\cite{Xie2016} address the formal verification of the HDFS reading and writing operations using CSP and the PAT model checker.
For instance, they formally modeled the reading and writing operations for the HDFS based on the formalized components proposed in~\cite{Su2009}.
Moreover, they verified some of the HDFS properties including the deadlock-freeness, minimal distance scheme, mutual exclusion, write-once scheme and robustness. While this approach allows to detect unexpected traces generating errors and verify data consistency in the HDFS, a limitation of this work is that it only models the reading and writing operations for just one file system and requires to investigate the validity of these operations for distributed files as in HDFS.
In \cite{Towards-Reddy2013}, Reddy \emph{et al.} propose an approach to model Hadoop's system using the PAT model checker. They used CSP to model Hadoop software components including the ``NameNode", ``DataNode", task scheduler and cluster setup. They identified the benefits of some properties like data locality, deadlock-freeness and non-termination among others and proved the correctness of these properties. However, their proposed model is evaluated on a small workload, and none of the properties is verified to check the performance of the Hadoop scheduler.
%Although theorem provers are widely used to check the correctness and reliability of several distributed systems,
%To the best of our knowledge, we only find one theorem prover-based study that verifies the actual running code of MapReduce applications.
Kosuke \emph{et al.}~\cite{OnoCoq2011} used the proof assistant Coq to write an abstract computation model of MapReduce. They modeled the mapper and reducer as Coq functions and proved that the implementation of the mapper and reducer satisfies the specification of some applications such as WordCount~\cite{Wordcount2015}.
The authors present an abstracted model of MapReduce, where many details are not included
%hidden\Foutse{are you saying that they didn't provided all details?} 
such as task assignment or resources allocation. These issues can affect the performance of applications running on Hadoop.
\vspace{-10pt} 

%% file: conclusion.tex
\section{Conclusion and Future Work}
\label{conclusion}
\vspace{-5pt}
Given the dynamic behavior of the cloud environment, the verification of the Hadoop scheduler has become an open challenge especially with many Hadoop-nodes being deployed to accommodate the increasing number of demands.
Existing approaches such as performance simulation and analytical modeling are not able to ascertain a complete verification of the Hadoop scheduler because of the wide range of constraints involved in Hadoop.
\Mbarka{In this paper, we presented a novel methodology that combines and integrates simulation and model checking techniques to perform a formal analysis of Hadoop schedulers. Particularly, we studied the feasibility of integrating model checking techniques and simulation to help in formally verifying some of the functional scheduling properties. So, we formally verified some of the most important scheduling properties on Hadoop including the schedulability, resources-deadlock freeness, and fairness properties, and analyzed their impact on the failure rates.
The ultimate goal of this work is to be able to propose possible scheduling strategies to reduce the number of failed tasks and improve the overall cluster performance and practitioner better assign and configure resources to mitigate the failures that a Hadoop scheduler may experience while simulating the Hadoop workload.}
We used CSP to model Hadoop schedulers, and the PAT model checker to verify the mentioned properties.
To show the practical usefulness of our work, we applied our approach on the scheduler of OpenCloud, a real Hadoop-based cluster.
%Our aim is to verify the three selected properties and check their impact on the tasks' failures.
The obtained results show that our approach is able to formally verify the selected properties and that such analysis can help identify up to 78\% of failures before they occur in the field.\\ %the failures when compared to the real-execution simulation traces of the OpenCloud scheduler.\\
\indent An important direction of future work is to verify other properties that can impact the failures rate in Hadoop, like the resource assignment, load balancing, and modeling the internal recovery mechanisms of Hadoop. Another direction can be the use of our work to automatically generate scheduling guidelines to improve the performance of the Hadoop framework and reduce the failures rate.
Finally, the failure analysis approach and findings of this paper can be extended and evaluated on Spark~\cite{SparkOnline}, which has become one of the key
cluster-computing framework that can be running on Hadoop. In fact, one can easily adapt the proposed methodology according to the architecture of Spark.